\documentclass[twocolumn,aps,pre,showpacs,nobalancelastpage,amssymb]{revtex4-1}
 
\usepackage{graphicx}
\usepackage{amsmath}
\usepackage{amssymb}
\usepackage{bm}
\usepackage{amsfonts}
\usepackage[usenames,dvipsnames]{xcolor}
\usepackage{appendix}
\usepackage[hidelinks]{hyperref}

\usepackage{color}

\begin{document}
\graphicspath{{figures/}}

\title{Ultrasensitive hysteretic force sensing with parametric nonlinear oscillators}
 
\author{Luca Papariello, Oded Zilberberg, Alexander Eichler, and R. Chitra}
\affiliation{Department of Physics, ETH Zurich, 8093 Z{\"u}rich, Switzerland.}

 
\begin{abstract}

We propose a novel method for linear detection of weak forces using parametrically driven nonlinear resonators. 
The method  is based on a peculiar feature in the response  of the resonator to a near resonant periodic external force.
This  feature stems from a complex interplay between the parametric drive, external force and nonlinearities. 
For weak parametric drive, the response exhibits the standard Duffing-like single jump hysteresis. 
For stronger drive amplitudes, we find a qualitatively new  double jump hysteresis which arises from stable solutions generated 
by the cubic Duffing nonlinearity. 
The additional jump exists only if the external force is present and the frequency at which it occurs depends linearly on the 
amplitude of the external force, permitting a straightforward ultrasensitive detection of weak forces. With state-of-the-art
nanomechanical resonators, our scheme should permit force detection in the atto-newton range.

\end{abstract}
 
 
\maketitle


\section{Introduction}

Research on nonlinear resonators started over a century ago, motivated by observations in electrodynamics and 
mechanics~\cite{nayfeh2008}. The fact that novel features are still discovered in nonlinear resonators today bears witness to 
their great complexity and variety. 
Nonlinear resonators manifest themselves in  many modern fields of physics, e.g. quantum electrical circuits, cold atoms, 
levitated nanoparticles, and nanoelectromechanical systems (NEMS)~\cite{DykmanBook}. 
They are intimately related to state-of-the-art metrology platforms used for measurements of weak 
external forces corresponding to single charges, spins, or mass on the atomic 
scale~\cite{RugarSpin, YangMass, PoggioNano, AlexNature}.

Interestingly, many of these modern resonators allow the study of individual modes whose nonlinearities can be tailored or tuned 
in situ and on which theoretical concepts, both classical and quantum, can be tested~\cite{Optomech}. 
One such concept is parametric resonance, where the frequency of the linear oscillator is modulated in 
time~\cite{PhysRevLett.67.699}. The parametrically driven oscillator boasts a fascinating stability diagram  called 
``Arnold's tongues'' delineating  zones where the oscillator is stable from those where it is exponentially unstable, as a function 
of its natural frequency and parametric driving strength~\cite{mclachlan1951theory}. 
In the stable regime, parametric resonance can be used to amplify signals and squeeze 
noise~\cite{Caves1981, PhysRevLett.67.699, Lehnert2007}, design mechanical logic circuits~\cite{Mahboob}, or generate 
quantum entanglement~\cite{Wallraff, Szorkovszky}. In the unstable regime, the resonator is 
driven to a large and stable response, which can be used for mechanical information storage~\cite{Yamaguchi} or signal 
amplification through bifurcation topology~\cite{Karabalin}.
 
Nonlinearities become important as resonators scale down~\cite{PostmaAPL}.
This can be attributed to geometric effects, external potentials, dissipation, or even feedback cooling used to control the resonator.  
Nonlinear effects strongly restrict the dynamic range  within which the system operates linearly, even making it vanishingly small 
in NEMS, and limit the scope for applications.   
However, recent works focus on directly using nonlinearities to improve the sensitivity of parametrically amplified 
detectors~\cite{ZhangNL, Karabalin, VillanuevaNonLin}. 
For instance, though quartic (Duffing) nonlinearities stabilize the parametric oscillator, it retains a ``memory'' of the underlying
instability tongue structure in its frequency dependent response~\cite{RandNonLin}.
The precision measurement of this lobe~\cite{ZhangNL, RandNonLin} then provides a very robust and stable way of 
detecting masses~\cite{ZhangNL}. 
Still, the utility of nonlinearities in parametrically driven oscillators for sensing of  external forces remains relatively unexplored.

In this work we obtain a solution for the response of an externally driven nonlinear parametric resonator  below and beyond the 
instability threshold.
The response  features an unexpected double hysteresis whose position depends sensitively and linearly on the amplitude of the
applied external force. Using recent experiments as examples, we predict how the double hysteresis should manifest, and we
propose a method to use it for the detection of weak forces. Importantly, the force sensor we propose has a linear dependency
of signal on measured force even though it is based on a nonlinear mechanical resonator.

The article is structured as follows. In Sec.~\ref{model}, we detail the model describing a general nonlinear parametric oscillator
model.  Section~\ref{method} is dedicated to a short description of the perturbative averaging method used for the analysis of the
model used to obtain a closed equation for the steady-state positional response.  
Based on our results for the response, we present our method for hysteretic force detection in Sec.~\ref{results}.  
In the concluding section, we discuss the application of our force detection scheme to different experimental systems.  
Discussion of the limiting cases and known results are relegated to the Appendices.

\section{Model}
\label{model}
The equation of motion governing the dynamics of a parametrically driven nonlinear oscillator of mass $m$ subject to a periodic
external force is
\begin{equation} \label{eq:FullEOM}
m \ddot{x} + m \omega_0^2 \left( 1 - \lambda \cos  \omega_p t  \right) x + 
\gamma \dot{x} + \alpha x^3 + \eta x^2 \dot{x} = F_0 \cos  \omega_f t  ,
\end{equation}
where $\omega_0$ is the unperturbed frequency of the linear oscillator, $\gamma$ the linear damping and
$\lambda$ and  $\omega_p$  denote the strength and frequency of the  parametric drive.
Parametric resonance occurs whenever the parametric drive frequency satisfies the condition $\omega_p = 2 \omega_0 / n$, 
where $n$ is an integer which labels the instability zones. The effects of parametric driving are most pronounced for $n \sim 1$. 
The nonlinearities are described by the Duffing parameter $\alpha$ characterizing the quartic contribution to the oscillator potential,
and  $\eta$  the strength of the nonlinear feedback cooling or nonlinear damping that is present in generic experimental 
setups~\cite{AlexNat, JanFeedback}.
Though nonlinearities  stabilize the regions of instability~\cite{RandNonLin, nayfeh2008}, the nonlinear  parametric resonator 
retains a precise memory of the instability regions of the parametric linear oscillator (see Appendix~\ref{nonlinpar}).  

The term on the right hand side  of \eqref{eq:FullEOM} refers to a periodic external force of strength $F_0$ and frequency 
$\omega_f$.   
Equation~\eqref{eq:FullEOM} generically describes the physics of resonators realized in a wide range of experimental
setups. Though a vast literature exists on the solutions to this equation in various regimes~\cite{LifshitzCross, nayfeh2008},
surprisingly, the impact of a periodically modulated external force on the full nonlinear problem has not been studied in great detail.
In the following we consider a positive Duffing parameter $\alpha$. Our methodology and results can be straightforwardly extended 
to the case of negative Duffing coefficients (as will be discussed later).
 
The main focus of this work involves studying the response of a parametric resonator to an external force, $F_0 \neq 0$, in the
nonlinear regime. 
Bifurcations arise in this problem which essentially change the nature of the associated response.  
Equation~\eqref{eq:FullEOM} is a non-autonomous, inhomogeneous and nonlinear differential equation that
does not permit an analytic solution for generic parameters. In typical experiments, the focus is on the first parametric resonance of 
the system, i.e. operating around twice the bare frequency of the undriven oscillator $\omega_p \approx 2 \omega_0$ while the
frequency of the external drive is $\omega_f \approx \omega_0$. As we will show, approximate analytic solutions to the frequency 
dependent response can be obtained in these experimentally relevant parameter regimes.

\section{Response function}
\label{method}
To analyze the equation of motion [Eq.~\eqref{eq:FullEOM}] we use the perturbative averaging 
method~\cite{guckenheimer1990nonlinear}, which replaces the full time dependent equation by time independent averaged 
equations of motion. 
Before that, we redefine time and space in Eq.~\eqref{eq:FullEOM} according to $\tau = \omega_0 t$ and 
$z = x \sqrt{\alpha / m \omega_0^2}$. This leads to the dimensionless equation of motion
\begin{equation} \label{eq:EOMdimLess}
\ddot{z} + \bar{\gamma} \dot{z} + z^3 + \bar{\eta} z^2 \dot{z} + \left( 1 - \lambda \cos 2 \Omega \tau \right) z = 
\bar{F}_0 \cos \Omega \tau ,
\end{equation}
\newline
where the dimensionless parameters are defined as $\bar{\gamma} \equiv \gamma / m \omega_0 = 1 / Q$, 
$\bar{\eta} \equiv \eta \omega_0 / \alpha$, $\Omega \equiv \omega / \omega_0$, and
$\bar{F}_0 \equiv ( F_0 / \omega_0^3 ) \sqrt{\alpha / m^3}$.
Using this method the frequency region around the first instability lobe is parametrized by setting $\omega_p = 2 \omega$, with 
$\omega \approx \omega_0$. Additionally, we introduce a detuning parameter $\sigma = 1 - \Omega^2$. 
Furthermore, the frequency of the external drive is locked at half the value of the parametric pump frequency $\omega_f = \omega$. 
With the parameters so defined, Eq.~\eqref{eq:FullEOM} can be recast as a pair of first order equations 
\begin{align} 
y & = \dot{z}\,,\\
\dot{y} + \omega^2 z &= f(z, y, t) \, ,
\label{eq:FullEOMy2}
\end{align}
with
\begin{align} \label{eq:f}
f(z, y, t) & = -\sigma z - \bar{\gamma} y - z^3 - \bar{\eta} z^2 y + 
\lambda \cos (2 \Omega t) z \nonumber \\
& \quad \, + \bar{F_0} \cos (\Omega t + \vartheta) \,,
\end{align}
where $\vartheta$ denotes the relative phase between the two drives.  
Note that, in order for the present perturbative method to be valid, the detuning
$\sigma$, linear and nonlinear damping $\bar{\gamma}$ and $\bar{\eta}$, as well as the driving strengths 
$\lambda$ and $\bar{F_0}$ have to be small.
The next step consists in bringing Eq.~\eqref{eq:FullEOMy2} into the so-called standard
form for averaging, i.e., into a form $\dot{z} = \epsilon f(z, y, t)$ with 
$0 < \epsilon \ll 1$. This is accomplished using the van der Pol transformation 
\cite{guckenheimer1990nonlinear} to  variables  $U$ and $V$, 
\begin{equation} \label{eq:Transf}
\left[ \begin{array}{c}
	z \\ y
	\end{array} \right] = 
	\begin{bmatrix}
	\cos \Omega t & - \sin \Omega t \\
	- \Omega \sin \Omega t & - \Omega \cos \Omega t
	\end{bmatrix} 
	\left[ \begin{array}{c}
	U \\ V
	\end{array} \right] .
\end{equation} 
Substituting \eqref{eq:Transf} in \eqref{eq:FullEOMy2} and  averaging over the time period $T = 2 \pi / \Omega$, we obtain 
the equations for the slow flow variables $u=\overline{U}$ and $v=\overline{V}$, which correspond to time-averaged $U$ and $V$ 
over the time cycle:
\begin{widetext}
\begin{align}
\dot{u} & = - \frac{1}{2 \Omega} \left[ \bar{\gamma} \Omega u + 
v \left( \sigma + \frac{\lambda}{2} \right) 
+ \frac{3}{4} (u^2 + v^2) v + \Omega \frac{\bar{\eta}}{4} (u^2 + v^2) u - \bar{F_0} \sin \vartheta \right] \,,
\label{eq:SlowU} \\
\dot{v} & = - \frac{1}{2 \Omega} \left[ \bar{\gamma} \Omega v 
+ u \left( - \sigma + \frac{\lambda}{2} \right) 
- \frac{3}{4} (u^2 + v^2) u + \Omega \frac{\bar{\eta}}{4} (u^2 + v^2) v + \bar{F_0} \cos \vartheta \right] \,.
\label{eq:SlowV}
\end{align} 
\end{widetext}
Despite the perturbative nature of the averaging method, it is valid for a surprisingly large
range of values of the drive amplitude $\lambda$ as well as for substantial detuning $\Omega$~\cite{Batista2007}. 

The coupled slow flow Eqs.~\eqref{eq:SlowU} and \eqref{eq:SlowV} remain  analytically insolvable. 
However, from the perspective of measurements, one only needs to know the
frequency response of the oscillator $\bar{|X|} = (u^2 + v^2)^{1/2}$.  
This is a property of the steady-state and does not require knowledge of transients.
Consequently, in the steady-state, we set ${\dot u} = {\dot v}=0$ in Eqs.~\eqref{eq:SlowU} and \eqref{eq:SlowV}, and we 
find that  the response $|X|^2$ satisfies the following polynomial equation 
\begin{widetext}
\begin{align} \label{eq:implX}
\bar{|X|}^2 & \left[ \left({\bar \gamma} \Omega + \frac{\bar \eta}{4} \Omega \bar{|X|}^2 \right)^2 - \left( \frac{{\lambda}}{2} \right)^2 + 
\left( \sigma + \frac{3}{4} \bar{|X|}^2\right)^2 \right]^2 = \nonumber \\
& = {\bar F}_0^2 \left[ \left( {\bar\gamma} \Omega + \frac{\bar \eta}{4} \Omega \bar{|X|}^2  \right)^2 
+ \left( \frac{{\lambda}}{2} \right)^2 + \left( \sigma + \frac{3}{4} \bar{|X|}^2 \right)^2 + 
{\lambda} \left( \sigma + \frac{3}{4} \bar{|X|}^2\right) \cos 2 \vartheta
+ \lambda \left( {\bar\gamma} \Omega + \frac{\bar \eta}{4} \Omega \bar{|X|}^2 \right) \sin 2 \vartheta \right] \, .
\end{align} 
\end{widetext}
Equation~\eqref{eq:implX} determines the response in a finite frequency interval $\omega$ around $\omega_0$.
Obtaining the response for arbitrary $\omega$ requires a non-perturbative approach or the retention of higher order corrections.
\begin{figure}[t] 
  \begin{center}
	\includegraphics[scale=1]{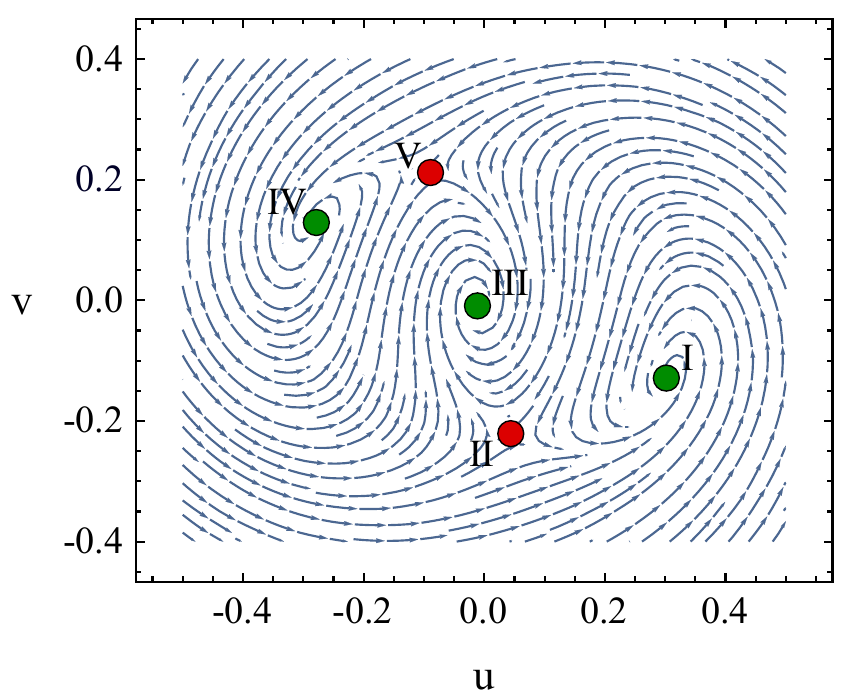}
	\caption{Trajectories of the  nonlinear parametric resonator in slow variables $u$ and $v$. The parameters chosen correspond to
	the unstable regime of the unforced linear parametric resonator:
	$\lambda = 5 \times 10^{-2}$, 
	${\bar F}_0 = 1 \times 10^{-3}$, ${\bar \gamma} = 1 \times 10^{-2}$, ${\bar \eta} = 3 \times 10^{-1}$,
        $\vartheta = 0$ and $\Omega = 1.03$. The green circles denote stable solutions and red unstable ones.}
	\label{fig:stream}
  \end{center}
\end{figure}

\section{Results}
\label{results}

\subsection{Response}
As will be shown below, the interplay between the periodic external force, parametric drive and nonlinearities leads to two 
qualitative different behaviors for the response depending on the position in parameter space. 
The solutions to the fifth order polynomial [Eq.~\eqref{eq:implX}]  can be stable or unstable.
The  stabilities  can be  directly inferred from the basins of attraction for this equation, plotted in Fig.~\ref{fig:stream}.
We find that typically one has three stable solutions (I, III and IV) marked by green dots and two unstable solutions (II and V)
denoted by the red dots. 
\begin{figure}[t] 
  \begin{center}
	\includegraphics[scale=1]{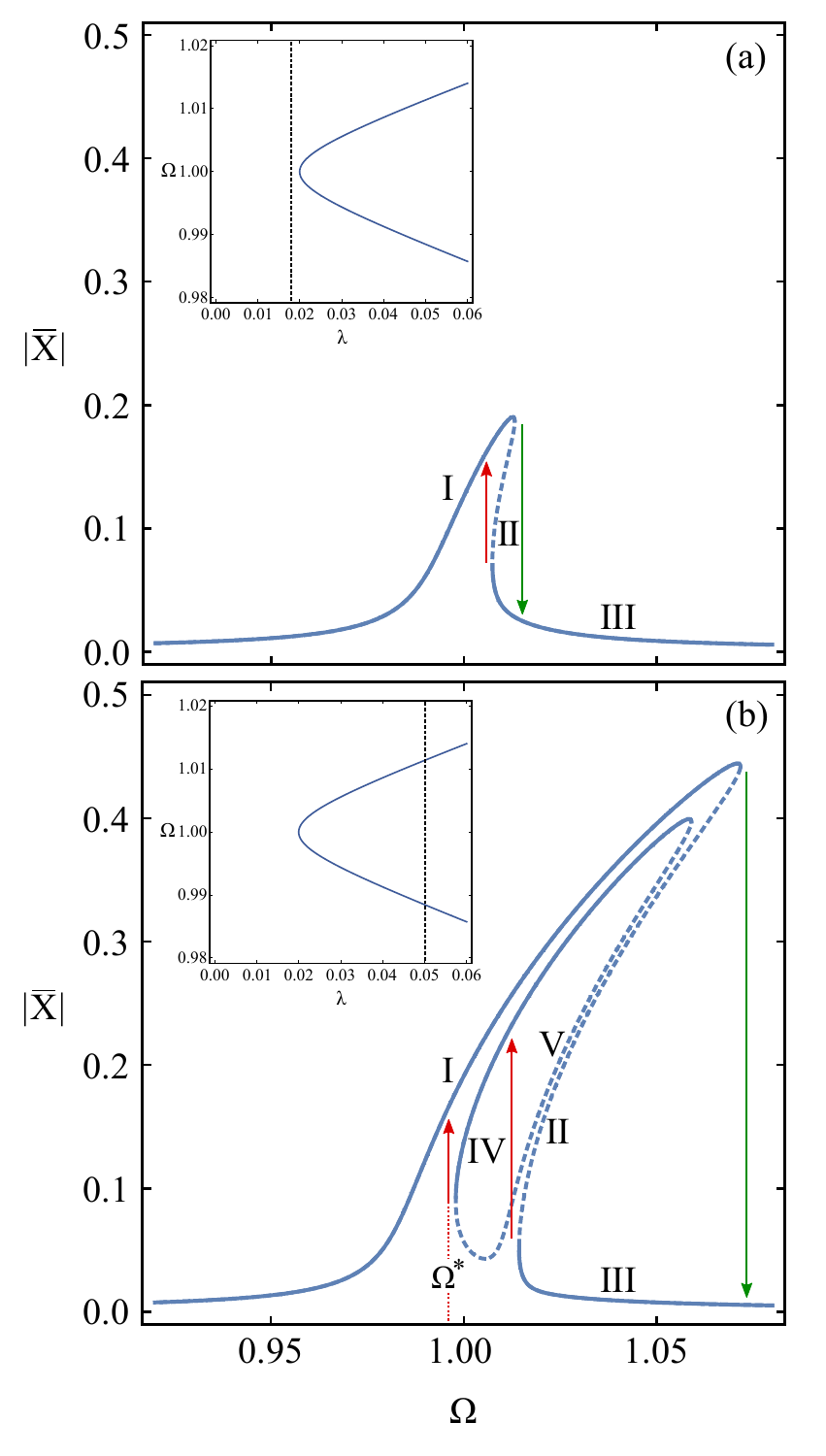}
	\caption{Typical frequency responses of the system described by Eq.~\eqref{eq:FullEOM}.
	The insets show the first instability region (Arnold tongue) of the unforced linear parametric resonator 
	($\alpha = \eta = F_0 = 0$)  (continuous lines) and the chosen values of $\lambda$ (dashed lines).
        (a) $\lambda = 1.8 \times 10^{-2}$ below the instability threshold (dashed line in inset);
	(b)  $\lambda = 5 \times 10^{-2}$ above the instability threshold (dashed line in inset). 
	Stable branches are indicated by whole lines, while unstable branches by dashed lines.
	The parameters  $\bar{F}_0 = 1 \times 10^{-3}$, 
        $\bar{\gamma} = 1 \times 10^{-2}$, $\bar{\eta} = 3 \times 10^{-1}$ and
        $\vartheta = 0$ are the same in both (a) and (b).}
	\label{fig:1}
  \end{center}
\end{figure}

For small amplitudes of the parametric drive $\lambda$ pertaining to the stable regime [see inset in Fig.~\ref{fig:1}(a)], the response  
shown in Fig.~\ref{fig:1}(a) is dominated by the external force and resembles that of the Duffing 
oscillator~\cite{guckenheimer1990nonlinear}. Here the stable solutions I and IV become degenerate and the
response has  two stable branches (I and III)  and one degenerate unstable branch (II).
As $\lambda$ increases and one crosses over to the unstable regime of the underlying linear oscillator 
[see inset in Fig.~\ref{fig:1}(b)], the degeneracies of both stable and unstable solutions are broken corresponding to the three 
stable attractors and two saddle points shown in Fig.~\ref{fig:stream}.  
This  generates  a qualitatively different response as shown in Fig.~\ref{fig:1}(b), with an enhanced Duffing-type response 
encompassing an island-like structure. This is due to a complex interplay between the cubic nonlinearity, the external force 
and the parametric drive.  
We reiterate that this response cannot be obtained without the periodic external force.

As $F_0$ increases, the island is raised and shifted to larger frequencies. A sufficiently strong $F_0$ wipes out  the internal 
island and the resulting frequency response is external force dominated and appears to be Duffing-like.
In the limit $F_0 \rightarrow 0$, we recover the response shown in Appendix~\ref{nonlinpar},  where I and IV (II and V) 
coalesce to a single stable (unstable) branch. The presence of $F_0$ thus leads to a splitting
of the stable (unstable) branch into two stable (unstable) branches. 

We now analyze the dependence of this novel response on the various tunable parameters in the system.
The driving strength $\lambda$  strongly affects  both the maximal amplitude of the response 
as well as the frequency at which the intermediate stable branch originates. 
As $\lambda$ increases,  the intermediate branch  dips further towards lower frequencies  though the  maximal response increases.
Linear damping $\gamma$, on the other hand simply   shifts the stability boundaries of the linear parametric oscillator away from the 
$\lambda = 0$ axis [see inset in Fig.~\ref{fig:1}(a)].
As a result, for the response, it plays a role akin to the inverse of the driving strength $\lambda$, i.e. the larger the damping, 
the smaller the response and  the origin of the intermediate  branch is pushed to higher frequencies.
For sufficiently large damping $\gamma$, one enters the parameter region where the linear oscillator is
stable and we recover the typical response of Fig.~\ref{fig:1}(a).
Importantly, nonlinear damping $\eta$  caps the response when $\omega$ increases, but it preserves the  intermediate stable 
branch and the island-like structure.

\subsection{Force Detection}
We will now show that the amplitude of the near resonant periodic external force can be directly extracted from the qualitatively 
new response discussed earlier. The presence of stable and unstable branches in the  response is expected to
lead to hysteretic behavior during upward and downward sweeps of the frequency $\omega$ across $\omega_0$. 
Consider the response for weak parametric driving plotted in Fig.~\ref{fig:1}(a). For upward sweeps of the frequency across 
$\omega_0$, the response will gradually increase along branch I all the way to the maximal value where it hits the
upper bifurcation and will then abruptly drop to the value of the lower stable branch III [green arrow in  Fig.~\ref{fig:1}(a)]. 
For downward sweeps, the response slowly increases along branch III and then jumps abruptly to the stable branch I (red arrow). 
This is very similar to the standard  Duffing-like hysteresis  seen in many systems both in the presence and absence of an external 
force~\cite{guckenheimer1990nonlinear, nayfeh2008}. The sizes of the hysteretic jumps depend on many parameters, including 
$F_0$. It is highly nontrivial to extract the amplitude of the force from this hysteresis curve.

For $\lambda$ in the unstable regime of the linear oscillator [cf. Fig.~\ref{fig:1}(b)],  the presence of additional branches in the 
response leads to a new kind of hysteresis curve. For upward sweeps across the
resonance frequency, the response will gradually increase all the way along branch I to the maximal value where it hits the
upper bifurcation and will then abruptly drop to the value of the lower stable branch III.  For downward sweeps,
the response will increase very slowly across branch III until it hits the first bifurcation
where it will abruptly jump to the stable branch IV of the island. It will then decrease further until it hits another bifurcation
of the island at  a frequency $\Omega^*$ where it will jump to the  stable branch I. In short, the presence of stable solutions in the
island results in two consecutive jumps in the downward sweeps.

The hysteretic jumps expected for the two response functions in the stable and unstable regimes are indicated in 
Fig.~\ref{fig:1}. Figure~\ref{fig:1}(b) shows a double jump hysteresis whereas Fig.~\ref{fig:1}(a) shows the
standard single jump hysteresis.
The second jump in Fig.~\ref{fig:1}(b)  is a direct manifestation of the intermediate stable branch discussed above and exists 
only when the amplitude of the external force $F_0$ is nonzero. The second jump is lost for high values of $\eta$ as the island shifts 
to higher frequencies. This feature  provides a promising new method to detect weak forces.
The force $F_0$  can be extracted either from the magnitude of the second jump or from the frequency $\Omega^{*}$ at which 
it occurs.

\begin{figure}[!h] 
  \begin{center}
	\includegraphics[scale=1]{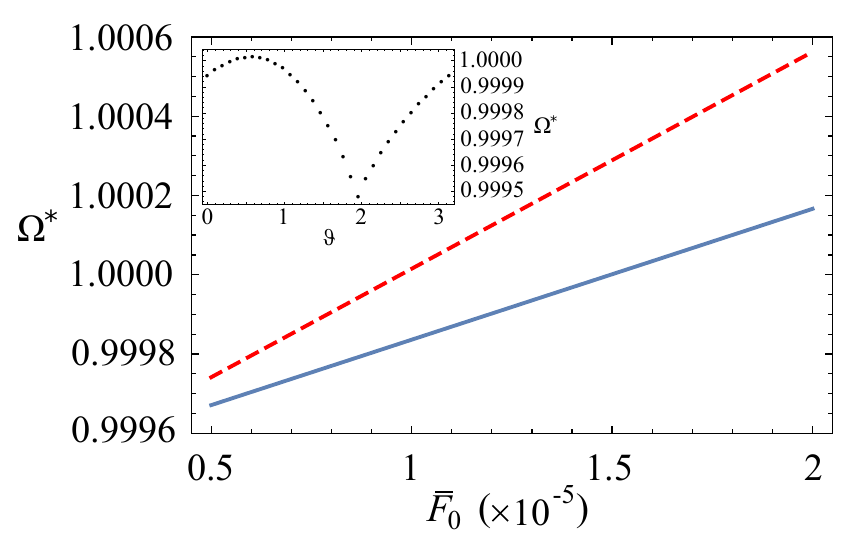}
	\caption{Jump frequency $\Omega^*$ as a function of the strength of the external force for $\vartheta = \pi / 4$ (red dashed line) 
	and averaged over a uniformly distributed phase $\vartheta$ (blue solid line).
	The parameters are given by $\lambda = 0.016$, $\bar{\gamma} = 10^{-3}$,
	$\bar{\alpha} = 7 \times 10^{-3}$,  $\bar{\eta} = 5 \times 10^{-3}$. 
	The inset shows the phase dependence of the jump frequency $\Omega^*$ for a fixed value of the external force 
	($\bar{F}_0 = 1 \times 10^{-5}$).}
	\label{fig:3}
  \end{center}
\end{figure}

\begin{figure}[!h] 
  \begin{center}
	\includegraphics[scale=1]{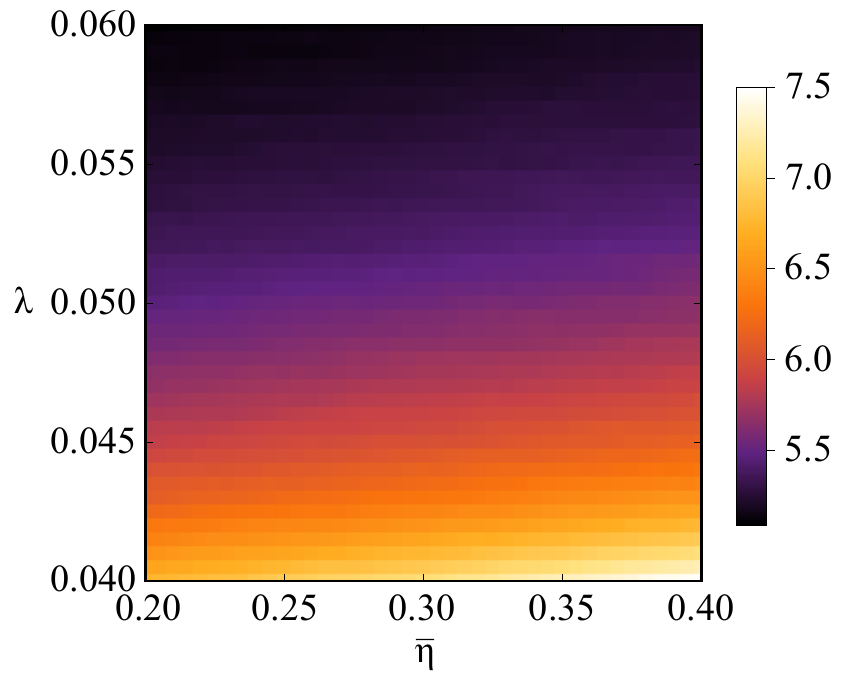}
	\caption{Sensitivity $\bar{\kappa}$ as a function of the strength of the parametric drive $\lambda$ and 
	nonlinear damping $\bar{\eta}$. The force range $\bar{F_0}$ is from $5 \times 10^{-4}$ to $1 \times 10^{-3}$, and  
	$\bar{\gamma} = 10^{-2}$.   
	The other parameters are kept fixed.}
	\label{fig:sens}
  \end{center}
\end{figure}

We find that $\Omega^*$ depends linearly on $F_0$ for a wide range of forces, allowing for a new and simple force detection 
scheme (see Fig.~\ref{fig:3}). 
The slope of $\Omega^*$ versus ${\bar F}_0$ ($\Omega^* = \omega^* / \omega_0$) defines a dimensionless sensitivity 
${\bar {\kappa}}$ which can be translated into physical units through the relation 
$\kappa = \frac{{\bar {\kappa}}}{\omega_0^2} \sqrt{ \frac{\vert \alpha \vert}{m^3} }$. The jump frequency, and thus the sensitivity, 
also depend on the relative phase between the periodic drive and the external force, as shown in the inset of Fig.~\ref{fig:3}. 
In the following, we consider the two cases that will be most relevant for experiments. On the one hand, if the phase $\vartheta$ of
$F_0$ is stable and can be controlled, one can reach the maximum sensitivity $\kappa_{\textrm{max}}$ that corresponds to 
$\vartheta \sim \pi / 4$ (red dashed line in Fig.~\ref{fig:3}). On the other hand, if the phase of $F_0$ is 
fluctuating, one effectively obtains a phase-averaged measurement with sensitivity $\kappa_{\textrm{mean}}$ (blue solid line in 
Fig.~\ref{fig:3}). In Fig.~\ref{fig:sens}, we plot the phase averaged dimensionless sensitivity of the device 
(${\bar {\kappa}_{\textrm{mean}}}$) as a function of $\lambda$ and ${\bar {\eta}}$. It is worth noting that as long as 
the parametric drive $\lambda$ is beyond the instability threshold, the sensitivity \textit{increases} with decreasing $\lambda$. 
We present values for both $\kappa_{\textrm{max}}$ and $\kappa_{\textrm{mean}}$ for typical experimental systems in the 
following section.

We note that a similar double jump hysteresis is  expected for a system with negative Duffing parameter $\alpha$ where the 
response tilts towards the left (spring softening)~\cite{nayfeh2008}. In this case, $\Omega^*$ decreases linearly with increasing 
$F_0$ but the sensitivity, given by the magnitude of the slope, is expected to be the same as that for positive $\alpha$.  
In other words, regardless of the sign of $\alpha$, a direct measurement of the hysteresis curve in the nonlinear regime of the 
parametrically driven resonator permits a straightforward extraction of the amplitude of the external force. 

Importantly, from an experimental perspective, one needs a nonlinear oscillator with well characterized Duffing nonlinearity 
and with tunable parametric modulation as well as nonlinear feedback cooling. 
The latter is particularly useful in  generating a sizable second hysteretic jump.
The device should first be calibrated, i.e. its sensitivity $\kappa$ should be obtained via a series of measurements of 
$\Omega^*$ for different values of known force amplitudes  $F_0$.  
Once the sensitivity is known, the device can be used to measure the amplitude of an unknown external force. 

\section{Discussion}

We now discuss the magnitudes of the forces that can be detected via the double hysteresis scheme. We consider an external 
force to be in principle detectable when the frequency shift of the second hysteresis is larger than the frequency noise present in 
the system, that is, if
\begin{equation} \label{force}
\kappa F_0 \geq \sigma_f,
\end{equation}
where $\kappa$ is the sensitivity of the device in physical units of angular frequency per force and we use $\sigma_f$ to denote the 
total (angular) frequency noise expected during a measurement. The minimum detectable force is then given by 
$F_{\textrm{min}} = \sigma_f / \kappa$. 
  
We present estimates for the range of forces which can be detected with two different resonators.
We first consider a laser-trapped nanoparticle in high vacuum~\cite{JanPRL} with a very high quality factor and a negative 
Duffing coefficient $\alpha$. This system allows for a wide manipulation of the system parameters with small thermal noise.
The system  parameters are: $m \approx 3 \times 10^{-18}$ kg, 
$\omega_0 \approx 2\pi \times 1.25 \times 10^5$ s$^{-1}$, $Q \approx 10^{8}$ (controlled through the air pressure) and 
$\vert \alpha \vert \approx  1.8 \times 10^{7}$ kg m$^{-2}$s$^{-2}$.
The nonlinear damping due to feedback cooling can be tuned in a range around $\eta \approx 14$ kg m$^{-2}$s$^{-1}$ and the
amplitude of the  parametric drive we use is $\lambda = 10^{-4}$, which is well inside the available modulation range.
Calculating $\bar{\kappa}$ from solutions of Eq.~\eqref{eq:implX} and then transforming into physical units, we obtain 
$\kappa_{\textrm{mean}} = 4 \times 10^{19}$ Hz/N and $\kappa_{\textrm{max}} = 5.6 \times 10^{19}$ Hz/N. 
For a sweep duration of typically a few seconds, the frequency noise can be expected to be in the range of 
$2\pi$ kHz in units of angular frequency~\cite{JanFeedback}, which gives a minimum detectable force of about $110$ aN and 
$160$ aN for $\kappa_{\textrm{max}}$ and $\kappa_{\textrm{mean}}$, respectively. Please note that the frequency noise used 
here is largely dominated by laser intensity noise and could in principle be decreased substantially.

The lightest nanomechanical resonators available today are made of individual carbon nanotubes. These resonators have 
pronounced nonlinearities and can be driven parametrically with high modulation depth~\cite{AlexNanoLett}. 
Typical parameters are~\cite{AlexNat}: $m \approx 10^{-20}$ kg, $\omega_0 \approx 2\pi \times 5 \times 10^7$ s$^{-1}$, 
$Q \approx 10^{3}$, $\eta \approx 10^{3}$ kg m$^{-2}$s$^{-1}$, 
$\vert \alpha \vert \approx  4 \times 10^{11}$ kg m$^{-2}$s$^{-2}$ and $\lambda = 2.5 \times 10^{-3}$. 
With these parameters, we get $\kappa_{\textrm{mean}} = 4.5 \times 10^{20}$ Hz/N and 
$\kappa_{\textrm{max}} = 7 \times 10^{20}$ Hz/N. From the linewidth of the frequency sweep in Fig.~4 of Ref.~\cite{AlexNat}, 
we estimate an upper bound for the frequency noise of $2\pi \times 5$ kHz in units of angular frequency, which result in minimum
detectable forces of 45 aN and 70 aN for $\kappa_{\textrm{max}}$ and $\kappa_{\textrm{mean}}$, respectively. 
The quality factor we use here is quite conservative. Values of up to $Q = 5\times 10^6$ have been measured more 
recently~\cite{Moser}. The same study also demonstrated significantly reduced frequency noise. 
However, it is not clear how the device will behave when driven into the nonlinear regime.

We expect weak  thermal fluctuations to  broaden the response  and modify the size of the hysteretic jumps, but  leave 
$\Omega^*$  effectively unchanged.  
As a result,  thermal noise will not have any qualitative impact on our detection scheme for devices with very high Q factors.  
Generically,  we expect the force to be detectable
as long as the second jump is visible  above the background noise. This should hold true as long as the system parameters 
as well as noise are such that one avoids activation of degenerate states or higher energy states.  
Note, the combination of driving, nonlinearities and noise could lead to phenomena similar to stochastic resonance in the
present context, but the study of these aspects is beyond the scope of the present work.
 
To conclude,  we have presented a new paradigm for  sensitive detection of forces using nonlinear parametric resonators. 
Though based on the nonlinear dynamics of the resonator,   our measurement scheme is inherently linear.
NEMS with relatively large Duffing nonlinearity $\alpha$ are  good candidates for our force detection scheme. 
For state-of-the-art devices, our scheme might allow the detection of forces in the $10 - 100$\,aN range.  
Furthermore, the high sensitivities  associated with our detection scheme can potentially be exploited in the context of
techniques such as nano-MRI aiming at great spatial resolution~\cite{Sidles, Degen}.


\acknowledgements

We thank J. Gieseler  for  numerous helpful discussions and comments on our work. 
We would also like to thank L. Novotny and E. Hebestreit for fruitful discussions.  We acknowledge financial support from the 
Swiss National Science Foundation.


\appendix

\section{Linear parametric oscillator}
\label{lin}
\begin{figure}[t] 
  \begin{center}
	\includegraphics[scale=0.9]{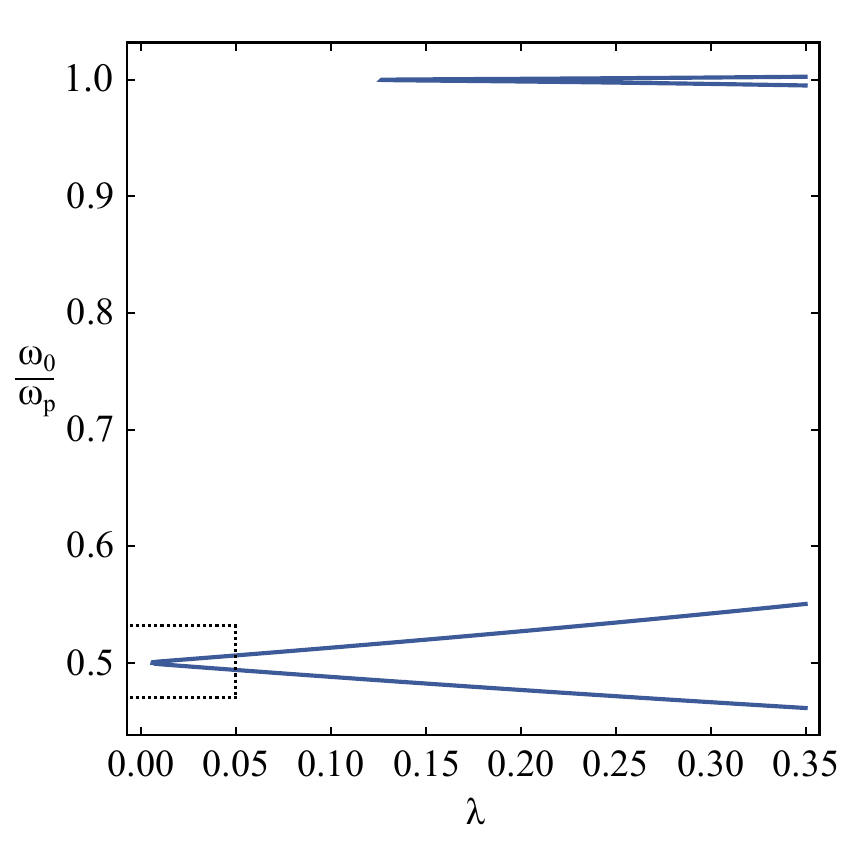}
	\caption{First two instability regions (Arnold tongues) of the parametrically driven oscillator (continuous lines). 
	The dashed box depicts the parameter region addressed in this work [cf. insets in Fig.~\ref{fig:1}].}
	\label{fig:Arnold}
  \end{center}
\end{figure}

The response of the linear parametric oscillator in the presence of a periodic external force has been analyzed in 
Ref.~\cite{PhysRevLett.67.699}. In Eq.~\eqref{eq:FullEOM} we set $\alpha = \eta = 0$. In this case, we obtain the
stability diagram with Arnold tongues shown in Fig.~\ref{fig:Arnold}.
For the first instability lobe, this response can easily be calculated from the slow flow equations [cf. Eqs.~\eqref{eq:SlowU} and
\eqref{eq:SlowV}] and has the form
\begin{equation}
|\bar{X}| = \frac{\sqrt{ (\bar{\gamma} \Omega)^2 + \sigma^2 + \left( \frac{\lambda}{2} \right)^2 + \lambda ( \sigma \cos 2 \vartheta + 
\bar{\gamma} \Omega \sin 2 \vartheta ) }}{ (\bar{\gamma} \Omega)^2 + \sigma^2 - \left( \frac{\lambda}{2} \right)^2 } ,
\end{equation}
where $\bar{\gamma} \equiv \gamma / m \omega_0$ and $\Omega \equiv \omega / \omega_0$. Here we have chosen  
$\omega_p = 2 \omega$ and $\omega_f = \omega$ with $\omega \approx \omega_0$, and we introduced the detuning parameter 
$\sigma = 1 - \Omega^2$.

\begin{figure}[t] 
  \begin{center}
	\includegraphics[scale=1]{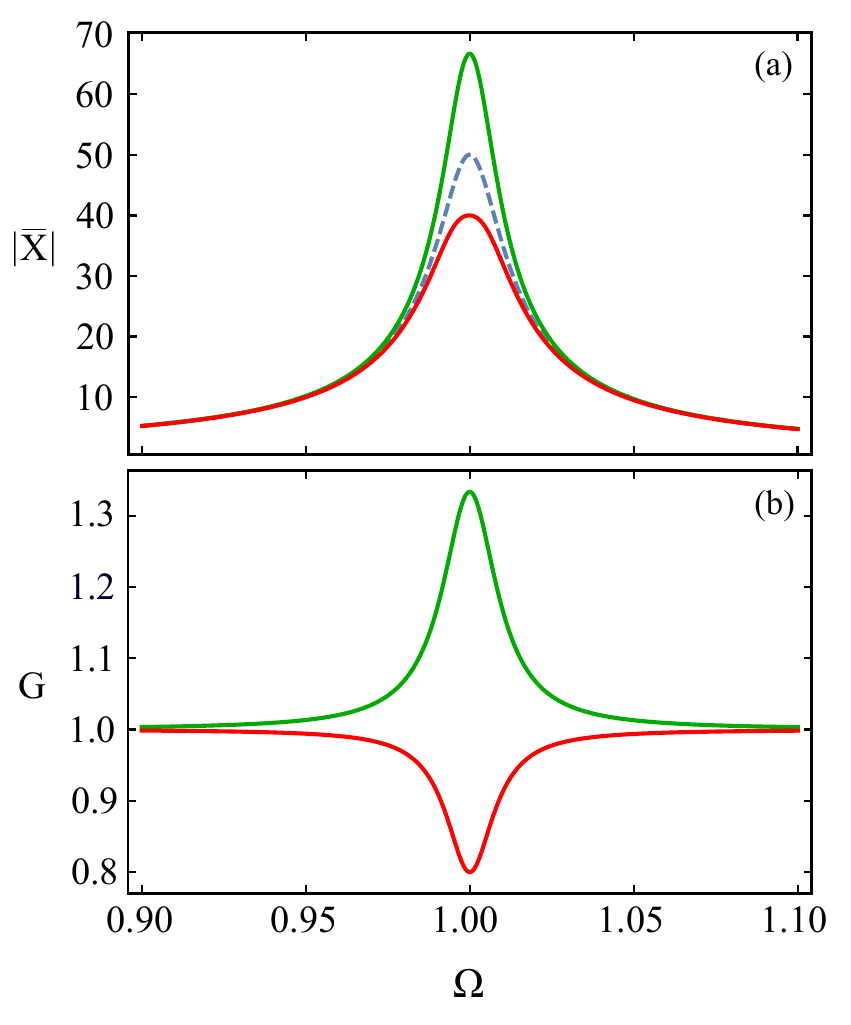}
	\caption{(a) Typical frequency response for the usual harmonic oscillator (dashed line) and the parametric oscillator 
	(continuous lines).
	For the red line $\vartheta = 3 \pi /4 $ and the response is suppressed, while for the green line $\vartheta = \pi /4 $ and the
	response is enhanced. In (b) we show the gain for the same two phases.}
	\label{fig:5_3}
  \end{center}
\end{figure}

The typical response for different regimes are plotted in Fig.~\ref{fig:5_3}(a).  Here we see that parametric driving enhances 
or reduces the response depending on the relative phase between direct and parametric drives. For instance for 
$\vartheta = \pi / 4$ (resp. $\vartheta = 3 \pi / 4$) we have a remarkable increase (resp. decrease) of the gain 
[see Fig.~\ref{fig:5_3}(b)]. Gain is here defined as $G = |\bar{X}|_{\lambda \neq 0} / |\bar{X}|_{\lambda = 0}$.

\section{Homogeneous non-linear parametric oscillator}
\label{nonlinpar}
We now discuss the response of the nonlinear parametric oscillator in the absence of an external driving force.  
This will help clarify the effect of nonlinearities, both Duffing type as well as feedback cooling, on the shape of the
response curves.
In the absence of an external force $F_0 = 0$, the steady-state response has a trivial solution $|\bar{X}| = 0$ as 
well as non-trivial solutions satisfying the equation
\begin{equation} \label{eq:impNoF}
 \left({\bar \gamma} \Omega + \frac{\bar \eta}{4} \Omega \bar{|X|}^2 \right)^2 - \left( \frac{{\lambda}}{2} \right)^2 + 
\left( \sigma + \frac{3}{4} \bar{|X|}^2\right)^2 = 0 \,.
\end{equation}

\subsection{Zero feedback cooling ($\eta=0$)}
 
The nontrivial solutions for $\alpha \neq 0$ and $\eta = 0$ are given by 
\begin{equation} \label{eq:FreqResp1}
|\bar{X}|^2 = \frac{4}{3} \left(
- \sigma \pm \sqrt{ \left( \frac{\lambda}{2} \right)^2 - (\bar{\gamma} \Omega)^2 } \right) \,.
\end{equation}

The presence  of the Duffing nonlinearity  effectively stabilizes the parametric oscillator and allows us to explore the previous
unstable region of parameter space.  The oscillator displacement does not increase  exponentially
but saturates to a fixed amplitude.
The  frequency-response of such a system (cf. Ref.~\cite{landau1976mechanics}) is plotted in Fig.~\ref{fig:7_2}. It
is characterized by three distinct regions with a different number of solutions each. In the first zone below A
there is a single stable solution. The second zone between A and B has
a high-amplitude stable solution and a zero-amplitude unstable one. The third zone beyond B has two stable solutions: a
zero-amplitude and a high-amplitude one, as well as an unstable solution between the two stable branches. 
The extent of the second region, which is delimited by the occurrence of pitchfork bifurcations~\cite{RandNonLin},  
is determined by the following equation $(\bar{\gamma} \Omega)^2 = (\lambda / 2)^2 - \sigma^2$. This corresponds exactly
to the equation for the first instability tongue~\cite{Batista2007}.  
The positive Duffing coefficient  $\alpha$ results in a rightward tilt of the response, reflecting the 
effective hardening of the spring constant. Note that a negative Duffing term would result in a tilt towards the left,
reflecting the softening of the spring constant.  This response has been measured in torsional MEMS~\cite{Stambaugh}.
%

\subsection{Feedback cooling ($\eta \neq 0$)}

\begin{figure}[t] 
  \begin{center}
	\includegraphics[scale=0.9]{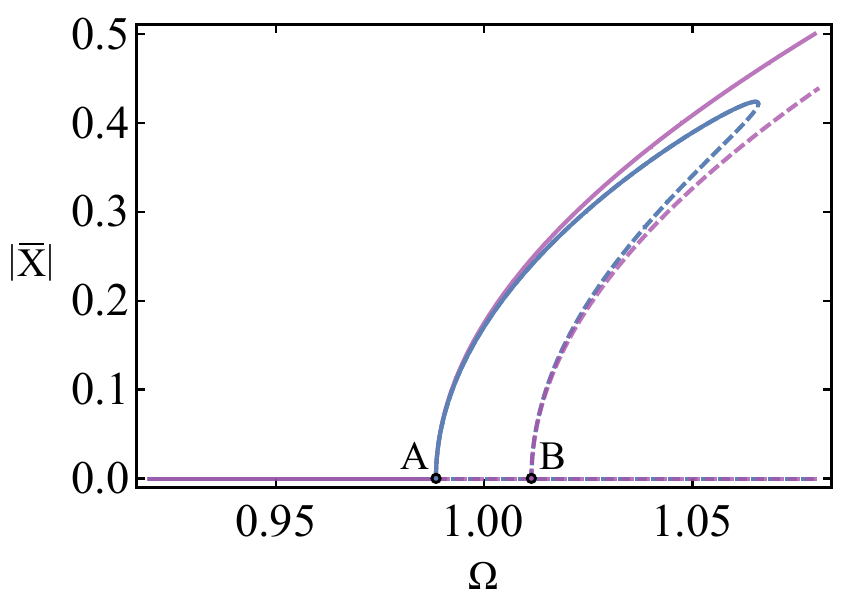}
	\caption{Typical frequency response for an unforced nonlinear parametric oscillator with nonlinear damping 
	$\eta \neq 0$ (blue line) and without nonlinear damping $\eta = 0$ (purple line). }
	\label{fig:7_2}
  \end{center}
\end{figure}

It is straightforward to assess the effect of 
nonlinear damping ($\eta \neq 0$) on the above response.
The resulting response (cf. Ref.~\cite{LifshitzCross}), which is shown is Fig.~\ref{fig:7_2}, is qualitatively similar to  
the one obtained for $\eta=0$. Nonlinear damping does not affect the bifurcations discussed earlier, but it principally limits the
growth of the response as the frequency $\omega$ increases.  

The detection of the width AB of the second region, which corresponds to tracing out the width of the Arnold tongue, was proposed as a way 
to do high precision mass sensing~\cite{ZhangNL}. High precision is expected because of the sharp changes in the response amplitude
at the boundaries of parametric resonance.
 
In both cases, the presence of stable and unstable solutions and a Duffing-like response is expected to
lead to hysteretic behavior
during  upward and downward sweeps of the frequency $\omega$ across $\omega_0$. For upward sweeps across the
resonance frequency, the response will gradually increase all the way to the maximal value where it hits the
upper bifurcation and will then abruptly drop to the value of the zero-amplitude stable branch.  For downward sweeps,
the response corresponds to the zero-amplitude stable branch until it hits the bifurcation point $B$,
where it will abruptly jump to the outer stable branch of the response.

\bibliographystyle{apsrev}
\bibliography{references}

\begin{thebibliography}{32}
\expandafter\ifx\csname natexlab\endcsname\relax\def\natexlab#1{#1}\fi
\expandafter\ifx\csname bibnamefont\endcsname\relax
  \def\bibnamefont#1{#1}\fi
\expandafter\ifx\csname bibfnamefont\endcsname\relax
  \def\bibfnamefont#1{#1}\fi
\expandafter\ifx\csname citenamefont\endcsname\relax
  \def\citenamefont#1{#1}\fi
\expandafter\ifx\csname url\endcsname\relax
  \def\url#1{\texttt{#1}}\fi
\expandafter\ifx\csname urlprefix\endcsname\relax\def\urlprefix{URL }\fi
\providecommand{\bibinfo}[2]{#2}
\providecommand{\eprint}[2][]{\url{#2}}

\bibitem[{\citenamefont{Nayfeh and Mook}(2008)}]{nayfeh2008}
\bibinfo{author}{\bibfnamefont{A.~H.} \bibnamefont{Nayfeh}} \bibnamefont{and}
  \bibinfo{author}{\bibfnamefont{D.~T.} \bibnamefont{Mook}},
  \emph{\bibinfo{title}{Nonlinear Oscillations}}, Physics textbook
  (\bibinfo{publisher}{Wiley}, \bibinfo{year}{2008}).

\bibitem[{\citenamefont{Dykman}(2012)}]{DykmanBook}
\bibinfo{author}{\bibfnamefont{M.}~\bibnamefont{Dykman}},
  \emph{\bibinfo{title}{Fluctuating Nonlinear Oscillators}}
  (\bibinfo{publisher}{Oxford University Press}, \bibinfo{year}{2012}).

\bibitem[{\citenamefont{Rugar et~al.}(2004)\citenamefont{Rugar, Budakian,
  Mamin, and Chui}}]{RugarSpin}
\bibinfo{author}{\bibfnamefont{D.}~\bibnamefont{Rugar}},
  \bibinfo{author}{\bibfnamefont{R.}~\bibnamefont{Budakian}},
  \bibinfo{author}{\bibfnamefont{H.~J.} \bibnamefont{Mamin}}, \bibnamefont{and}
  \bibinfo{author}{\bibfnamefont{B.~W.} \bibnamefont{Chui}},
  \bibinfo{journal}{Nature} \textbf{\bibinfo{volume}{430}},
  \bibinfo{pages}{329} (\bibinfo{year}{2004}).

\bibitem[{\citenamefont{Yang et~al.}(2006)\citenamefont{Yang, Callegari, Feng,
  Ekinci, and Roukes}}]{YangMass}
\bibinfo{author}{\bibfnamefont{Y.~T.} \bibnamefont{Yang}},
  \bibinfo{author}{\bibfnamefont{C.}~\bibnamefont{Callegari}},
  \bibinfo{author}{\bibfnamefont{X.~L.} \bibnamefont{Feng}},
  \bibinfo{author}{\bibfnamefont{K.~L.} \bibnamefont{Ekinci}},
  \bibnamefont{and} \bibinfo{author}{\bibfnamefont{M.~L.}
  \bibnamefont{Roukes}}, \bibinfo{journal}{Nano Letters}
  \textbf{\bibinfo{volume}{6}}, \bibinfo{pages}{583} (\bibinfo{year}{2006}).

\bibitem[{\citenamefont{Poggio}(2013)}]{PoggioNano}
\bibinfo{author}{\bibfnamefont{M.}~\bibnamefont{Poggio}}, \bibinfo{journal}{Nat
  Nano} \textbf{\bibinfo{volume}{8}}, \bibinfo{pages}{482}
  (\bibinfo{year}{2013}).

\bibitem[{\citenamefont{Chaste et~al.}(2012)\citenamefont{Chaste, Eichler,
  Moser, Ceballos, Rurali, and Bachtold}}]{AlexNature}
\bibinfo{author}{\bibfnamefont{J.}~\bibnamefont{Chaste}},
  \bibinfo{author}{\bibfnamefont{A.}~\bibnamefont{Eichler}},
  \bibinfo{author}{\bibfnamefont{J.}~\bibnamefont{Moser}},
  \bibinfo{author}{\bibfnamefont{G.}~\bibnamefont{Ceballos}},
  \bibinfo{author}{\bibfnamefont{R.}~\bibnamefont{Rurali}}, \bibnamefont{and}
  \bibinfo{author}{\bibfnamefont{A.}~\bibnamefont{Bachtold}},
  \bibinfo{journal}{Nat Nano} \textbf{\bibinfo{volume}{7}},
  \bibinfo{pages}{301} (\bibinfo{year}{2012}).

\bibitem[{\citenamefont{Aspelmeyer et~al.}(2014)\citenamefont{Aspelmeyer,
  Kippenberg, and Marquardt}}]{Optomech}
\bibinfo{author}{\bibfnamefont{M.}~\bibnamefont{Aspelmeyer}},
  \bibinfo{author}{\bibfnamefont{T.~J.} \bibnamefont{Kippenberg}},
  \bibnamefont{and}
  \bibinfo{author}{\bibfnamefont{F.}~\bibnamefont{Marquardt}},
  \bibinfo{journal}{Rev. Mod. Phys.} \textbf{\bibinfo{volume}{86}},
  \bibinfo{pages}{1391} (\bibinfo{year}{2014}).

\bibitem[{\citenamefont{Rugar and Gr\"utter}(1991)}]{PhysRevLett.67.699}
\bibinfo{author}{\bibfnamefont{D.}~\bibnamefont{Rugar}} \bibnamefont{and}
  \bibinfo{author}{\bibfnamefont{P.}~\bibnamefont{Gr\"utter}},
  \bibinfo{journal}{Phys. Rev. Lett.} \textbf{\bibinfo{volume}{67}},
  \bibinfo{pages}{699} (\bibinfo{year}{1991}).

\bibitem[{\citenamefont{McLachlan}(1951)}]{mclachlan1951theory}
\bibinfo{author}{\bibfnamefont{N.}~\bibnamefont{McLachlan}},
  \emph{\bibinfo{title}{Theory and application of Mathieu functions}}
  (\bibinfo{publisher}{Clarendon}, \bibinfo{year}{1951}).

\bibitem[{\citenamefont{Caves}(1981)}]{Caves1981}
\bibinfo{author}{\bibfnamefont{C.~M.} \bibnamefont{Caves}},
  \bibinfo{journal}{Phys. Rev. D} \textbf{\bibinfo{volume}{23}},
  \bibinfo{pages}{1693} (\bibinfo{year}{1981}).

\bibitem[{\citenamefont{Castellanos-Beltran and Lehnert}(2007)}]{Lehnert2007}
\bibinfo{author}{\bibfnamefont{M.~A.} \bibnamefont{Castellanos-Beltran}}
  \bibnamefont{and} \bibinfo{author}{\bibfnamefont{K.~W.}
  \bibnamefont{Lehnert}}, \bibinfo{journal}{Appl. Phys. Lett.}
  \textbf{\bibinfo{volume}{91}}, \bibinfo{pages}{083509}
  (\bibinfo{year}{2007}).

\bibitem[{\citenamefont{Mahboob et~al.}(2011)\citenamefont{Mahboob, Flurin,
  Nishiguchi, Fujiwara, and Yamaguchi}}]{Mahboob}
\bibinfo{author}{\bibfnamefont{I.}~\bibnamefont{Mahboob}},
  \bibinfo{author}{\bibfnamefont{E.}~\bibnamefont{Flurin}},
  \bibinfo{author}{\bibfnamefont{K.}~\bibnamefont{Nishiguchi}},
  \bibinfo{author}{\bibfnamefont{A.}~\bibnamefont{Fujiwara}}, \bibnamefont{and}
  \bibinfo{author}{\bibfnamefont{H.}~\bibnamefont{Yamaguchi}},
  \bibinfo{journal}{Nat. Commun.} \textbf{\bibinfo{volume}{2}},
  \bibinfo{pages}{198} (\bibinfo{year}{2011}).

\bibitem[{\citenamefont{Eichler et~al.}(2014)\citenamefont{Eichler, Salathe,
  Mlynek, Schmidt, and Wallraff}}]{Wallraff}
\bibinfo{author}{\bibfnamefont{C.}~\bibnamefont{Eichler}},
  \bibinfo{author}{\bibfnamefont{Y.}~\bibnamefont{Salathe}},
  \bibinfo{author}{\bibfnamefont{J.}~\bibnamefont{Mlynek}},
  \bibinfo{author}{\bibfnamefont{S.}~\bibnamefont{Schmidt}}, \bibnamefont{and}
  \bibinfo{author}{\bibfnamefont{A.}~\bibnamefont{Wallraff}},
  \bibinfo{journal}{Phys. Rev. Lett.} \textbf{\bibinfo{volume}{113}},
  \bibinfo{pages}{110502} (\bibinfo{year}{2014}).

\bibitem[{\citenamefont{Szorkovszky et~al.}(2014)\citenamefont{Szorkovszky,
  Clerk, Doherty, and Bowen}}]{Szorkovszky}
\bibinfo{author}{\bibfnamefont{A.}~\bibnamefont{Szorkovszky}},
  \bibinfo{author}{\bibfnamefont{A.~A.} \bibnamefont{Clerk}},
  \bibinfo{author}{\bibfnamefont{A.~C.} \bibnamefont{Doherty}},
  \bibnamefont{and} \bibinfo{author}{\bibfnamefont{W.~P.} \bibnamefont{Bowen}},
  \bibinfo{journal}{New J. Phys.} \textbf{\bibinfo{volume}{16}},
  \bibinfo{pages}{063043} (\bibinfo{year}{2014}).

\bibitem[{\citenamefont{Mahboob and Yamaguchi}(2008)}]{Yamaguchi}
\bibinfo{author}{\bibfnamefont{I.}~\bibnamefont{Mahboob}} \bibnamefont{and}
  \bibinfo{author}{\bibfnamefont{H.}~\bibnamefont{Yamaguchi}},
  \bibinfo{journal}{Nat Nano} \textbf{\bibinfo{volume}{3}},
  \bibinfo{pages}{275} (\bibinfo{year}{2008}).

\bibitem[{\citenamefont{Karabalin et~al.}(2011)\citenamefont{Karabalin,
  Lifshitz, Cross, Matheny, Masmanidis, and Roukes}}]{Karabalin}
\bibinfo{author}{\bibfnamefont{R.~B.} \bibnamefont{Karabalin}},
  \bibinfo{author}{\bibfnamefont{R.}~\bibnamefont{Lifshitz}},
  \bibinfo{author}{\bibfnamefont{M.~C.} \bibnamefont{Cross}},
  \bibinfo{author}{\bibfnamefont{M.~H.} \bibnamefont{Matheny}},
  \bibinfo{author}{\bibfnamefont{S.~C.} \bibnamefont{Masmanidis}},
  \bibnamefont{and} \bibinfo{author}{\bibfnamefont{M.~L.}
  \bibnamefont{Roukes}}, \bibinfo{journal}{Phys. Rev. Lett.}
  \textbf{\bibinfo{volume}{106}}, \bibinfo{pages}{094102}
  (\bibinfo{year}{2011}).

\bibitem[{\citenamefont{Postma et~al.}(2005)\citenamefont{Postma, Kozinsky,
  Husain, and Roukes}}]{PostmaAPL}
\bibinfo{author}{\bibfnamefont{H.~W.~C.} \bibnamefont{Postma}},
  \bibinfo{author}{\bibfnamefont{I.}~\bibnamefont{Kozinsky}},
  \bibinfo{author}{\bibfnamefont{A.}~\bibnamefont{Husain}}, \bibnamefont{and}
  \bibinfo{author}{\bibfnamefont{M.~L.} \bibnamefont{Roukes}},
  \bibinfo{journal}{Applied Physics Letters} \textbf{\bibinfo{volume}{86}},
  \bibinfo{eid}{223105} (\bibinfo{year}{2005}).

\bibitem[{\citenamefont{Zhang et~al.}(2002)\citenamefont{Zhang, Baskaran, and
  Turner}}]{ZhangNL}
\bibinfo{author}{\bibfnamefont{W.}~\bibnamefont{Zhang}},
  \bibinfo{author}{\bibfnamefont{R.}~\bibnamefont{Baskaran}}, \bibnamefont{and}
  \bibinfo{author}{\bibfnamefont{K.~L.} \bibnamefont{Turner}},
  \bibinfo{journal}{Sensors and Actuators A: Physical}
  \textbf{\bibinfo{volume}{102}}, \bibinfo{pages}{139} (\bibinfo{year}{2002}).

\bibitem[{\citenamefont{Villanueva et~al.}(2013)\citenamefont{Villanueva,
  Kenig, Karabalin, Matheny, Lifshitz, Cross, and Roukes}}]{VillanuevaNonLin}
\bibinfo{author}{\bibfnamefont{L.~G.} \bibnamefont{Villanueva}},
  \bibinfo{author}{\bibfnamefont{E.}~\bibnamefont{Kenig}},
  \bibinfo{author}{\bibfnamefont{R.~B.} \bibnamefont{Karabalin}},
  \bibinfo{author}{\bibfnamefont{M.~H.} \bibnamefont{Matheny}},
  \bibinfo{author}{\bibfnamefont{R.}~\bibnamefont{Lifshitz}},
  \bibinfo{author}{\bibfnamefont{M.~C.} \bibnamefont{Cross}}, \bibnamefont{and}
  \bibinfo{author}{\bibfnamefont{M.~L.} \bibnamefont{Roukes}},
  \bibinfo{journal}{Phys. Rev. Lett.} \textbf{\bibinfo{volume}{110}},
  \bibinfo{pages}{177208} (\bibinfo{year}{2013}).

\bibitem[{\citenamefont{Rand}(2000)}]{RandNonLin}
\bibinfo{author}{\bibfnamefont{R.~H.} \bibnamefont{Rand}}
  (\bibinfo{year}{2000}).

\bibitem[{\citenamefont{Eichler
  et~al.}(2011{\natexlab{a}})\citenamefont{Eichler, Moser, Chaste, Zdrojek,
  Wilson-Rae, and Bachtold}}]{AlexNat}
\bibinfo{author}{\bibfnamefont{A.}~\bibnamefont{Eichler}},
  \bibinfo{author}{\bibfnamefont{J.}~\bibnamefont{Moser}},
  \bibinfo{author}{\bibfnamefont{J.}~\bibnamefont{Chaste}},
  \bibinfo{author}{\bibfnamefont{M.}~\bibnamefont{Zdrojek}},
  \bibinfo{author}{\bibfnamefont{I.}~\bibnamefont{Wilson-Rae}},
  \bibnamefont{and} \bibinfo{author}{\bibfnamefont{A.}~\bibnamefont{Bachtold}},
  \bibinfo{journal}{Nat Nano} \textbf{\bibinfo{volume}{6}},
  \bibinfo{pages}{339} (\bibinfo{year}{2011}{\natexlab{a}}).

\bibitem[{\citenamefont{Gieseler et~al.}(2012)\citenamefont{Gieseler, Deutsch,
  Quidant, and Novotny}}]{JanFeedback}
\bibinfo{author}{\bibfnamefont{J.}~\bibnamefont{Gieseler}},
  \bibinfo{author}{\bibfnamefont{B.}~\bibnamefont{Deutsch}},
  \bibinfo{author}{\bibfnamefont{R.}~\bibnamefont{Quidant}}, \bibnamefont{and}
  \bibinfo{author}{\bibfnamefont{L.}~\bibnamefont{Novotny}},
  \bibinfo{journal}{Phys. Rev. Lett.} \textbf{\bibinfo{volume}{109}},
  \bibinfo{pages}{103603} (\bibinfo{year}{2012}).

\bibitem[{\citenamefont{Lifshitz}(2009)}]{LifshitzCross}
\bibinfo{author}{\bibfnamefont{M.~C.} \bibnamefont{Lifshitz},
  \bibfnamefont{R.~Cross}}, \emph{\bibinfo{title}{Nonlinear Dynamics of
  Nanomechanical and Micromechanical Resonators}}
  (\bibinfo{publisher}{Wiley-VCH}, \bibinfo{year}{2009}), pp.
  \bibinfo{pages}{1--52}.

\bibitem[{\citenamefont{Guckenheimer and
  Holmes}(1990)}]{guckenheimer1990nonlinear}
\bibinfo{author}{\bibfnamefont{J.}~\bibnamefont{Guckenheimer}}
  \bibnamefont{and} \bibinfo{author}{\bibfnamefont{P.}~\bibnamefont{Holmes}},
  \emph{\bibinfo{title}{Nonlinear oscillations, dynamical systems, and
  bifurcations of vector fields}}, Applied mathematical sciences
  (\bibinfo{publisher}{Springer-Verlag}, \bibinfo{year}{1990}).

\bibitem[{\citenamefont{Batista}(2011)}]{Batista2007}
\bibinfo{author}{\bibfnamefont{A.~A.} \bibnamefont{Batista}},
  \bibinfo{journal}{Journal of Statistical Mechanics: Theory and Experiment}
  \textbf{\bibinfo{volume}{2011}}, \bibinfo{pages}{P02007}
  (\bibinfo{year}{2011}).

\bibitem[{\citenamefont{Gieseler et~al.}(2014)\citenamefont{Gieseler,
  Spasenovi\ifmmode~\acute{c}\else \'{c}\fi{}, Novotny, and Quidant}}]{JanPRL}
\bibinfo{author}{\bibfnamefont{J.}~\bibnamefont{Gieseler}},
  \bibinfo{author}{\bibfnamefont{M.}~\bibnamefont{Spasenovi\ifmmode~\acute{c}\else
  \'{c}\fi{}}}, \bibinfo{author}{\bibfnamefont{L.}~\bibnamefont{Novotny}},
  \bibnamefont{and} \bibinfo{author}{\bibfnamefont{R.}~\bibnamefont{Quidant}},
  \bibinfo{journal}{Phys. Rev. Lett.} \textbf{\bibinfo{volume}{112}},
  \bibinfo{pages}{103603} (\bibinfo{year}{2014}).

\bibitem[{\citenamefont{Eichler
  et~al.}(2011{\natexlab{b}})\citenamefont{Eichler, Chaste, Moser, and
  Bachtold}}]{AlexNanoLett}
\bibinfo{author}{\bibfnamefont{A.}~\bibnamefont{Eichler}},
  \bibinfo{author}{\bibfnamefont{J.}~\bibnamefont{Chaste}},
  \bibinfo{author}{\bibfnamefont{J.}~\bibnamefont{Moser}}, \bibnamefont{and}
  \bibinfo{author}{\bibfnamefont{A.}~\bibnamefont{Bachtold}},
  \bibinfo{journal}{Nano Letters} \textbf{\bibinfo{volume}{11}},
  \bibinfo{pages}{2699} (\bibinfo{year}{2011}{\natexlab{b}}).

\bibitem[{\citenamefont{Moser et~al.}(2014)\citenamefont{Moser, Eichler,
  G{\"u}ttinger, Dykman, and Bachtold}}]{Moser}
\bibinfo{author}{\bibfnamefont{J.}~\bibnamefont{Moser}},
  \bibinfo{author}{\bibfnamefont{A.}~\bibnamefont{Eichler}},
  \bibinfo{author}{\bibfnamefont{J.}~\bibnamefont{G{\"u}ttinger}},
  \bibinfo{author}{\bibfnamefont{M.~I.} \bibnamefont{Dykman}},
  \bibnamefont{and} \bibinfo{author}{\bibfnamefont{A.}~\bibnamefont{Bachtold}},
  \bibinfo{journal}{Nat Nano} \textbf{\bibinfo{volume}{9}},
  \bibinfo{pages}{1007} (\bibinfo{year}{2014}).

\bibitem[{\citenamefont{Sidles}(1991)}]{Sidles}
\bibinfo{author}{\bibfnamefont{J.~A.} \bibnamefont{Sidles}},
  \bibinfo{journal}{Appl. Phys. Lett.} \textbf{\bibinfo{volume}{58}},
  \bibinfo{pages}{2854} (\bibinfo{year}{1991}).

\bibitem[{\citenamefont{Degen et~al.}(2009)\citenamefont{Degen, Poggio, Mamin,
  Rettner, and Rugar}}]{Degen}
\bibinfo{author}{\bibfnamefont{C.~L.} \bibnamefont{Degen}},
  \bibinfo{author}{\bibfnamefont{M.}~\bibnamefont{Poggio}},
  \bibinfo{author}{\bibfnamefont{H.~J.} \bibnamefont{Mamin}},
  \bibinfo{author}{\bibfnamefont{C.~T.} \bibnamefont{Rettner}},
  \bibnamefont{and} \bibinfo{author}{\bibfnamefont{D.}~\bibnamefont{Rugar}},
  \bibinfo{journal}{Proceedings of the National Academy of Sciences}
  \textbf{\bibinfo{volume}{106}}, \bibinfo{pages}{1313} (\bibinfo{year}{2009}).

\bibitem[{\citenamefont{Landau and Lifshitz}(1976)}]{landau1976mechanics}
\bibinfo{author}{\bibfnamefont{L.}~\bibnamefont{Landau}} \bibnamefont{and}
  \bibinfo{author}{\bibfnamefont{E.}~\bibnamefont{Lifshitz}},
  \emph{\bibinfo{title}{Mechanics}}, Butterworth-Heinemann
  (\bibinfo{year}{1976}).

\bibitem[{\citenamefont{Chan and Stambaugh}(2007)}]{Stambaugh}
\bibinfo{author}{\bibfnamefont{H.~B.} \bibnamefont{Chan}} \bibnamefont{and}
  \bibinfo{author}{\bibfnamefont{C.}~\bibnamefont{Stambaugh}},
  \bibinfo{journal}{Phys. Rev. Lett.} \textbf{\bibinfo{volume}{99}},
  \bibinfo{pages}{060601} (\bibinfo{year}{2007}).

\end{thebibliography}

\end{document}